\documentclass[12pt,preprint]{aastex}
\newcommand{\bq}{\begin{equation}}
\newcommand{\eq}{\end{equation}}
\newcommand{\h}{^h}
\newcommand{\m}{^m}
\newcommand{\s}{^s}
\newcommand{\dg}{^\circ}

\newcommand{\am}{'}
\newcommand{\3}{$_3$}
\newcommand{\as}{''}
\newcommand{\simgt}{\lower.5ex\hbox{$\; \buildrel > \over \sim \;$}}
\newcommand{\simlt}{\lower.5ex\hbox{$\; \buildrel < \over \sim \;$}}
\newcommand{\sol}{_\odot}
\newcommand{\kms}{km s$^{-1}$}
\newcommand{\gnu}{$_{\it{v}}$}
\newcommand{\gnusp}{$_{\it{v}}$ }
\newcommand{\hone}{H{\tiny \sc I} }

\begin{document}
\title{Exploring High-Velocity NH\3(6,6) Emission at the Center of our Galaxy}
\author{Jennifer L. Donovan\altaffilmark{1}, Robeson M. Herrnstein\altaffilmark{1}, and Paul T.P. Ho\altaffilmark{2,3}}
\altaffiltext{1}{Department of Astronomy, Columbia University, 550 West 120th St., Mail Code 5246, New York, NY 10027; jen@astro.columbia.edu, herrnstein@astro.columbia.edu}
\altaffiltext{2}{Harvard-Smithsonian Center for Astrophysics, 60 Garden St., Cambridge, MA 02138; pho@cfa.harvard.edu}
\altaffiltext{3}{Academia Sinica Institute of Astronomy and Astrophysics, Taipei}

\begin{abstract}
Using the NH\3 (6,6) transition, which samples dense  ($\sim 10^{5}$) molecular gas with an energy above ground of 412~K, we find hot gas at high velocities (--142 to --210 km~s$^{-1}$) associated with the central 2~pc of the Galactic center. This material may be either infalling gas due to shocks or tidal stripping, or possibly gas swept from the nuclear region. We identify two high-velocity features, which we call the Southern Runner and the Cap, and correlate these features with others detected in various molecular observations of the Galactic center. The characteristic linewidths of the Southern Runner and Cap, 10 -- 15 \kms, are similar to those of other hot Galactic center clouds. The estimated H$_{2}$ masses of these clouds are 4$\times 10^{3}$ M$\sol$ and 2$\times 10^{3}$ M$\sol$, consistent with the masses of the western streamer and northern ridge, NH\3 (6,6) emission features detected within the central 10~pc at lower velocities. Three possible explanations for this emission are discussed assuming that they lie at the Galactic center, including sweeping by the supernova remnant Sgr A East, infall and/or shock from the circumnuclear disk (CND), and stripping from the central rotating low-velocity NH\3 (6,6) cloud.\end{abstract}


\section{Introduction}
The environment around Sgr A*, the black hole at the center of our Galaxy with an inferred mass of 4 $\times$ 10$^{6}$ M$\sol$ (e.g., \citealt{Schodel03, Ghez05}), is a topic of interest to diverse fields. This laboratory for the testing of general relativity around a supermassive black hole is surrounded by massive stars (e.g., \citealt{Morris99}), sites of intense star formation (e.g., \citealt{Schodel05}), molecular clouds (e.g., \citealt{Pedlar89, Coil00, Zhao95}), and hot X-ray plasma (e.g., \citealt{Rockefeller04}), each interconnected in the overall picture of the Galactic center. Understanding separately the details of each of these components is a crucial task in itself, as well as an important step toward the comprehension of other galactic nuclei not located so conveniently nearby. Hot, dense molecular gas clouds are of particular interest, as they feed both star formation and accretion onto the black hole, two prominent topics of research for galaxies at all redshifts. 

Various tracers illuminate the numerous gaseous features at the Galactic center. The circumnuclear disk (CND), a rotating ring of molecular gas, neutral hydrogen, and dust that surrounds the central 2 pc of the Galaxy \citep{Gusten87, Becklin82, Liszt83}, has been studied in HCN, CO, \hone absorption, and in mid- and far-infrared emission. The inner, ionized edge of the negative velocity CND lobe forms the western arm of the spiral-shaped H{\tiny \sc II} region known as Sgr A West, or the minispiral; the two other ionized filamentary components are known as the bar and the northern arm. Sgr A West is best mapped in radio continuum \citep{Ekers83} or recombination lines such as H92$\alpha$ \citep{Roberts93}. The near edge of the synchrotron-emitting supernova remnant Sgr A East approaches Sgr A* and the CND from behind \citep{Pedlar89}, and according to recent Chandra observations, may have just passed by Sgr A* \citep{Maeda02}. 

The presence of high-velocity gas is evident across the central several parsecs (\citealt{Gusten81}, hereafter GD81; \citealt{Zhao95, Herrnstein02}); in particular, high negative velocity gas (HNVG) has been detected in absorption by several studies. GD81 first recognized the absorption of \hone and H$_2$CO in the velocity range of --160 to --210 km s$^{-1}$ and proposed its cause to be an outflow from the black hole. These observations were followed up in studies by \citet{Marr92} and \citet{Pauls93}, in which HCO$^+$ at --180 km s$^{-1}$ was detected in absorption against Sgr A West and H$_{2}$CO was detected over a range of velocities extending to --210 km s$^{-1}$ in absorption near Sgr A*. The relative positions of the absorption and continuum led \citet{Pauls93} to conclude that the HNVG was behind Sgr A* and, from this position, falling into the minispiral. 

\citet{Yusef93} imaged the \hone gas in the velocity range --160 to --211 km s$^{-1}$ first detected by GD81 and suggested IRS 16, a cluster of young He emission line stars within the central 2~pc, as the origin of the wind accelerating this high-velocity gas. The presence of young stars at the very center of the Galaxy is a subject of much current research, as these stars must either be able to form at small Galactic radii (fractions of a parsec) or migrate inward much more efficiently than is thought possible \citep{Ghez03}. In fact, \citet{Ghez03} illuminated this ``paradox of youth" issue with the detection of a massive (15 M$\sol$) star with an age of less than 10 Myr in an orbit with a semimajor axis of only 125.6 mas (5.0 $\times$ 10$^{-3}$ pc). \citet{Eisenhauer05} further found that over 90\% of the stars detected within 0.$\as$5 of Sgr A* are normal, main sequence B0-B9 stars that likely formed in their present locations. The existence of stars so young is also suggestive of the presence of molecular gas, perhaps left over after $\it{in}$ $\it{situ}$ star formation, and high velocity molecular gas such as has been observed for the past 25 years is quite possibly being accelerated by the winds and shocks from these young stars.

Further observations of such HNVG were performed by \citet{Liszt93}, as they mapped the same cloud as GD81 in CO emission at --190 km s$^{-1}$, although they found no evidence that the cloud was physically associated with Sgr A*. Instead, they suggested that it was associated with the expanding molecular ring \citep{Bania77} at least 150~pc in front of the black hole \citep{Liszt93}. Finally, \citet{Zhao95} detected OH in absorption toward Sgr A West at --180 km s$^{-1}$, but they concluded that this gas was closer to the Galactic center than to the gas of the expanding molecular ring.

Few high-velocity studies have been done, however, with NH\3, another important tracer of dense ($\sim10^5$~cm$^{-3}$) molecular gas at the Galactic center. The fractional population of each transition detected is dependent upon the temperature of the gas; for example, more (1,1) is detected at low temperatures than (6,6) though both are present. \citet{McGary01} and \citet{Herrnstein02} observed the central 10~pc in four rotation inversion transitions of the molecule across the velocity range --140 to +130 km s$^{-1}$. They concluded in \citet{Herrnstein05}, hereafter HH05, that the temperature structure at the Galactic center can be described by a two-temperature system where roughly one quarter of the molecular gas comprises a hot ($\approx$ 200~K) component, and the remaining gas is cool ($\approx$ 25~K), consistent with the temperature structure established by \citet{Huttemeister93} in a study of 36 NH\3 clouds at the Galactic center. Warm molecular clouds at the Galactic center, consistent with this hot component and ranging in temperature from 100 to 300~K, have been studied using pure-rotational molecular hydrogen lines by \citet{RodFer00, RodFer01}.

NH\3 detections are of particular interest in studying how the CND, and potentially Sgr A*, are fed, as the (3,3) transition of this molecule has been shown to trace the ``southern streamer''. This extension of the 20 \kms $\,$cloud has been suggested as interacting with the CND \citep{Okumura89, Ho91, Coil99, Coil00}, though absorption along the LOS makes the data difficult to interpret conclusively (HH05). The weakening of NH\3(3,3) emission as the streamer crosses the CND may indicate some interaction between this cloud and the nucleus (HH05). Two additional connections between the same GMC and the CND \citep{Coil99, Coil00} as well as one originating in the northern ridge \citep{McGary01} have also been shown to be detectable in NH\3 lines.  

Observations of NH\3 have shown important links between the CND and other molecular features at the Galactic center, but in order to get a more complete picture of the behavior of this species of molecular gas in the region -- especially in light of the presence of clouds at even higher velocities observed in absorption toward Sgr A*-- it is necessary to know whether NH\3 is present in more extreme velocity ranges. The observations described in this paper probe --142 to --210 km s$^{-1}$ with higher resolution than the previous ammonia observations by HH05, allowing for the detailed study of detected line profiles as well as the detection of narrow line features. We have chosen the NH\3 (6,6) line in order to determine whether the excitation, kinematics, and line widths traced by this transition can help to identify the origin and nature of the high negative velocity gas. In this paper, we describe our observations of NH\3(6,6) and the data reduction (\S 2), as well as our results (\S 3), and place them in the larger framework of other molecular observations and discuss their implications (\S 4). We summarize our results in \S 5, where we also describe necessary future work.

\section{Observations}

We observed the central 2~pc of the Galaxy (centered on Sgr A*, $\alpha_{2000}$=17$\h$45$\m$40.$\s$03, $\delta_{2000}$ =--29$\dg$00$\am$28.$\as$06) at 25.056025 GHz, the frequency of the NH\3 (6,6) transition, for one hour of integration time on source using the D north-C configuration of the Very Large Array\footnote{The National Radio Astronomy Observatory is a facility of the National Science Foundation operated under cooperative agreement by Associated Universities, Inc.} on 2003 February 2. The hybrid configurations of the VLA are useful for studies of such low declination sources. In the hybrid configuration, although the north-south baselines are physically three times longer than those east-west, they appear to be the same length when projected onto the plane perpendicular to the LOS, producing a circular beam in the resulting image. A phase calibrator was observed every fifteen minutes, roughly the coherence time of the atmosphere at K band. The 6.25 MHz bandwidth used was divided into 31 channels such that velocities from --140 km s$^{-1}$ to --210 km s$^{-1}$ were covered with a velocity resolution of 2.3 km s$^{-1}$. Channel 1 was omitted from the analysis due to the rolloff in the passband, resulting in velocity coverage from --142 km s$^{-1}$ to --210 km s$^{-1}$. Mindful of the fact that other dense Galactic center clouds have characteristic linewidths of 10 -- 15 km s$^{-1}$, this high velocity resolution provided the opportunity to study in detail the line profiles of detected clouds as well as to discover narrow-line features in the region at high negative velocity. 

The bandpass calibration was performed using the source 1229+020. This calibration is good to better than 1\% of the peak flux density of the continuum; therefore, the flux density of any feature detected at the position of Sgr A* (flux density $\sim$1~Jy) is accurate to 10 mJy. Continuum emission from Sgr A West has flux density levels $\simlt 0.25$ Jy, corresponding to, at most, an associated noise level of 2.5 mJy, which is less than the measured rms noise in our images (see \S 3).

The narrow channel widths (and the resultant narrow velocity window) also produced a marked lack of a set of completely line-free channels throughout the data cube, which became a concern for the continuum subtraction. We performed the continuum subtraction using three different sets of channels that were line-free for most of the data cube, and we consistently found the same features to be significant. For this reason, we believe that the subtraction is robust. The data presented here use the edge channels for the continuum subtraction. 

We used the Astronomical Image Processing System (AIPS) to reduce our data, performing self-calibration with both phase and amplitude using the 1~Jy point source, Sgr A*, and reducing the rms noise in the map by a factor of 3. To enhance our sensitivity to larger structures, we applied a Gaussian taper to the $\it{uv}$ data of 13.75 k$\lambda$ (15$\as$) at the 30$\%$ level, which resulted in a beam size of 10.8$\as$ $\times$ 8.6$\as$. The theoretical rms noise for this observation is $\sim$2~mJy, and our measured rms noise in a single channel ranges from 3 -- 4 mJy beam$^{-1}$; for this paper we quote the rms noise of our maps to be 4 mJy beam$^{-1}$. Therefore the noise from the bandpass calibration does not dominate (except for any feature whose position coincides with that of Sgr A*.) In order to eliminate negative sidelobes in our maps due to incomplete sampling of the $\it{uv}$ plane, we used clean boxes around the features that we will call the Southern Runner (centered at $\Delta\alpha$ = 10$\as$, $\Delta\delta$ = --65$\as$ relative to Sgr A*) and the Cap (centered at $\Delta\alpha$ = 35$\as$, $\Delta\delta$ = 35$\as$ relative to Sgr A*). The clouds centered at $\Delta\alpha$ = --30$\as$, $\Delta\delta$ = --40$\as$ and $\Delta\alpha$ = --15$\as$, $\Delta\delta$ = 70$\as$ (relative to Sgr A*) do not belong to these putative features and are not discussed in this work. The former is not as significant a detection as the Southern Runner and Cap, and the latter is located outside of our primary beam. Both are potentially interesting, however, and would be well-served by deeper and/or more spatially extended observations.

\section{Results}

The NH\3 (6,6) emission in each of our 30 channels from --142 to --210 km s$^{-1}$ is shown in Figure 1. The Southern Runner appears strongly ($>$3.5$\sigma$) in the --194, --196, and --198 km s$^{-1}$ channels, as does the Cap, though more tenuously. Emission from the clouds included in the Cap also appear in the --159, --161, --163, and --166 km s$^{-1}$ channels. We also detect faint emission from the southernmost lobe of the Southern Runner (spectrum H in Figure 3) outside of the primary beam in the --163, --166, --168, and --170 \kms channels. The significance of these features is best illustrated by integrating this emission over the entire velocity range, and this map is shown in Figure 2, in positive and negative contours of 4$\sigma$, 6.5$\sigma$, and 9$\sigma$, where 1$\sigma$ is mathematically determined to be 22.5 mJy beam$^{-1}$ km s$^{-1}$. 

Calculating $\sigma$, rather than determining it experimentally, is necessary in this case due to the method by which the integrated intensity, or moment zero, map was produced. In order to include only significant emission in this map, the moment zero map was made using the MOMNT task in AIPS with Boxcar smoothing over eight velocity channels and Boxcar smoothing with a kernel size of 9$\as$, clipping all flux with absolute value below 5~mJy in order to discriminate against noise features. As a result, the noise in the moment zero map is not Gaussian, nor is it consistent across the map (because it depends upon the number of channels included in the summation at each pixel.) We calculated $\sigma$ as follows: 

\begin{equation}
\sigma_{map} = \sigma_{ch} \Delta \rm{v} \sqrt{N_{ch}}
\end{equation}

\noindent where $\sigma_{ch}$ is the rms noise in one channel, $\Delta \rm{v}$ is the velocity resolution of one channel, and N$_{ch}$ is the typical number of velocity channels observed for any one feature. The rms noise in any one channel is 4 mJy beam$^{-1}$, and each feature covered six velocity channels on average, so our characteristic value for 1$\sigma$ is 22.5 mJy beam$^{-1}$ \kms.

The velocity-integrated NH\3 (6,6) is overlaid on a grayscale image of the minispiral from \citet{Roberts93} in Figure 2. Note that the location of the Cap coincides spatially with that of the trailing end of the east-west arc of the minispiral; the juxtaposition of these two features was the origin of the name for this particular feature. 

The negative features seen to the west of the Southern Runner and Cap in Figure 2 appear to be ``negative bowls", artifacts of the imaging process, present when the $\it{uv}$-coverage is missing short spacings and the emission is extended. They emerge at the same velocities as the emission features (see Figure 1) and have been reduced as much as possible without over-cleaning by drawing clean boxes around the positive features during the reduction process; before box cleaning, the negatives were symmetrically distributed around the positive features. To avoid over-cleaning, these features were only cleaned to the level of the noise in the surrounding channels (4 mJy beam$^{-1}$). (The reader is referred to Figure 1, which displays the consistency of the noise between box-cleaned and non-box-cleaned channels.) The clean boxes were then removed and the entire image was cleaned for typically another 900 iterations. 

Spectra taken using MIRIAD toward each emission feature, as well as their positions overlaid on a velocity-integrated map, are shown in Figures 3 and 4. Spectrum I has been included to show the level of the rms noise. Both the Southern Runner and the Cap exhibit emission at velocities of --160 and --195 \kms, which will be discussed further in \S 4, and three position-velocity diagrams have been included in Figure 4 to clarify this emission. The Southern Runner, shown in the top right panel, is most likely a coherent feature with a velocity gradient of 2 \kms arcsec$^{-1}$. The Cap, which is cut in two different directions and shown in the two lower panels of Figure 4, appears to consist of multiple clouds at different velocities. Interestingly, the two apparent north-south features (one marked by spectra A and B, the other by spectra C and D) each appear to exhibit both velocities. The position-velocity diagrams shown in Figure 4 display lower-level emission than the associated velocity-integrated map because they were made with an un-smoothed data cube in order to show low-level emission that may have been eliminated as noise in the creation of the moment zero map.

\subsection{Mass Estimation}

Estimating the masses of these two features, though inherently uncertain, is one way to compare this high negative velocity emission to other clouds in the region. Kinematic and spatial comparisons are made in the following section. Following the mass estimation technique outlined in HH05 and making similar assumptions, we find column densities for NH\3 of 

\begin{equation}
N_{NH_3} = 1.8  \times  10^{15} \rm{cm}^{-2} \Bigg( \frac{\Delta \rm{v}}{1.5 \times 10^{6} cm~s^{-1}} \Bigg) \Bigg(\frac{T_{b}}{1.5~K} \Bigg) 
\end{equation}

\noindent where $\Delta$v is the linewidth and T$_{b}$ is the brightness temperature. This equation ignores the hyperfine satellites, as 96.9\% of the emission is in the main line for this transition \citep{HoTownes83}. It also takes into account the calculation performed in \citet{Ho77}, which indicates that the fractional population of the NH\3(6,6) transition is $\approx$ 0.1 at 200~K. For 100-300~K, the range of temperatures of hot gas at the Galactic center \citep{Huttemeister93, RodFer00, RodFer01}, the same fractional population of NH\3(6,6) is valid at 300~K, and it drops by a factor of 3 at 100~K; for gas at the low end of this range of temperatures, the column density quoted above would at most increase by a factor of three (see \citet{RobinThesis} for a detailed discussion of the dependence of the fractional population on temperature.) Assuming a temperature of 200~K for the NH\3(6,6) emission detected would make this gas consistent with the hot component of the two-temperature structure described in HH05 and \citet{Huttemeister93}. For consistency with HH05, we also consider only the lowest 19 metastable states of NH\3 in this fractional population, as considering all transitions will at most increase our population estimate by a factor of two at 200~K. The NH\3 (6,6) transition occurs at a temperature of 412~K above ground.

T$_{b}$ is constrained by the observations since it can be obtained from the line profiles of our detected features \citep{Ho83} using 

\begin{equation}
T_{b} = \frac{\lambda^{2} \rm{r}^{2}}{2 k R^{2}} S_{\it{v}}
\end{equation}

\noindent where r is the distance to the source (8~kpc), R is the beam size, and S\gnusp is the integrated flux density in Janskys. For these observations, T$_{b}$ = 22.5~K (S\gnu), and as S\gnusp for our strongest detected spectral features ranges from 60 to 75 mJy, T$_{b}$ can be estimated to be 1.5~K. 

We convert the column density of NH\3 to solar masses of H$_{2}$ in the following manner:

\begin{equation}
M_{H_2} = 2.4 M\sol \Bigg( \frac{N_{NH_3}}{10^{15} \rm{cm}^{-2}} \Bigg) \Bigg( \frac{A} {1 \square \as} \Bigg) \Bigg( \frac{X(NH_3)}{10^{-8}} \Bigg)^{-1} 
\end{equation}

\noindent where A is the area of the cloud in square arcseconds. As in HH05, we assume that the X(NH\3) factor, or the abundance of NH\3 relative to H$_{2}$, is on the order 10$^{-8}$; note that Equation (6) of that paper should indicate that the assumed value of X(NH\3) was 10$^{-8}$. This factor has been observed in dark, star-forming clouds to be of order 10$^{-9}$ \citep{Benson83, Batrla84, Serabyn86}, but in clouds with warm temperatures, the NH\3 abundance is suspected to be increased by evaporation of the molecule off of dust grains \citep{Pauls83, Walmsley87}. A value of X(NH\3) $\sim$ 10$^{-8}$ is also consistent with the range of measurements made by comparing warm ammonia column densities \citep{Huttemeister93} with warm H$_{2}$ column densities, as shown in \citet{RodFer01}.

For the Southern Runner, the above column density and the spatial extent of the cloud (a total area of approximately 20$\as$ by 50$\as$) translate to a mass of 4$\times 10^{3}$ M$\sol$. For the combined clouds of the Cap, with a total spatial extent of approximately 20$\as$ by 20$\as$, the mass estimate is 2$\times 10^{3}$ M$\sol$. These mass estimates are dependent upon the assumptions of temperature and X(NH\3) that we have made, and therefore they should be taken as order of magnitude estimates. 

\section{Discussion}

In this paper, it is assumed that the observed NH\3 (6,6) emission lies in the central parsecs of the Galaxy. If the features were merely along the same line of sight, it becomes difficult to make consistent the relatively large velocity differences between the clouds with the small projected distances between them. In addition, the detection of the (6,6) transition implies a high level of excitation, which would be unexpected in a normal galactic molecular cloud. However, NH\3 (6,6) has been detected at a projected distance of 300~pc from the Galactic center with a similar column density and attributed to collisions with excited H$_{2}$ in the ``expanding molecular ring" \citep{Gardner87}, which is now believed to be the signature of gas dynamics in a barred galaxy \citep{Liszt78,Liszt80}. Molecular clouds at higher Galactic longitudes associated with the bar have been observed to exhibit large velocity gradients over small spatial distance, up to 200 \kms \space over 30 pc \citep{Liszt06}, of which the Souther Runner (with a gradient of 40 \kms \space over 0.6 pc) could be an extreme example. The extreme velocities of the Southern Runner and Cap at a Galactic longitude of essentially 0$\dg$ fall on the edge or just outside of the parallelogram-shaped $\it{lv}$ diagram which represents the gas dynamics of the bar as traced by various studies in CO and OH \citep{Liszt80,BitranThesis,Huttemeister98,Sawada04}. More observational data are necessary to rule out an association of the presented features with the EMR/bar, but in this paper it is assumed that the observed emission lies in close proximity to the Galactic center and not simply along the line of sight. 

These newly detected high negative velocity clouds must be placed in the context of other high- and low-velocity clouds in the region. This allows the determination of whether we have detected some rapidly moving component of a well-known cloud or a completely new population of hot, high-velocity clouds in the region. In either event, it has been shown that emission at this high-velocity does exist in the nuclear region, and more observations are necessary to determine the extent of such velocity coverage by NH\3 and other dense-gas-tracing molecules. 

The kinematic measurements and mass estimates of the high-velocity (presented in this paper) and low-velocity (presented in HH05) NH\3 (6,6) features discussed in this section are summarized in Table 1, as are the high negative velocity OH absorption and CO emission clouds described in the introduction and the NH\3 (6,6) shown to be part of the EMR/bar. Though the gas detected via these three tracers may be related, none of the latter clouds appear to sample the same population as the high negative velocity NH\3 (6,6) emission presented in this work. The gas detected in OH absorption by \citet{Zhao95} and the (spatially distant) NH\3 (6,6) emission detected by \citet{Gardner87} have much larger linewidths though comparable mass, and that detected in CO emission by \citet{Liszt93} appears to be an order of magnitude too massive compared to the high negative velocity NH\3 (6,6) presented here. In the following subsections, we discuss the relation of the high negative velocity NH\3 (6,6) to the major features in the region, and we introduce three different, alternative explanations for the high velocity emission. For a visual aid to the confusing geometry of the Galactic center clouds, the reader is referred to the schematic drawings of the region in Figures 14, 15, and 16 of HH05.

\subsection{Sgr A East}

The supernova explosion that created Sgr A East was a major source of disturbance for the molecular gas in the Galactic center region, and it is thought to be having a direct effect upon certain observed molecular features, such as the northern ridge and the western streamer (HH05). We suggest the possibility that the molecular gas presented here is being likewise affected. 

Figure 5 shows our velocity-integrated high negative velocity NH\3 (6,6) overlaid in blue contours on Sgr A East in color from \citet{Yusef87}. (The dynamic range necessary to see the level of emission from Sgr A East is extreme; the red contours display these levels over a dynamic range of 10$^{4}$.) From the positional overlay, it appears that the NH\3 (6,6) emission lies completely interior to the edge of Sgr A East (marked by the outermost red contour line.) It is possible that the Southern Runner could lie on the near face of the expanding remnant, much like a vein on a cantaloupe, with the majority of its motion along the LOS.

The feasibility of this explanation is dependent upon the energy of the supernova(e) that created Sgr A East, which has been a matter of some dispute. \citet{Mezger89} suggested that the SNR cleared out from the central parsecs much of the material in the 50 \kms GMC, indicating an energy $\geq$4 $\times 10^{52}$ ergs. In HH05, the authors refuted this idea by showing that the supposed connection between the SNR and the GMC, the ``northern ridge", was kinematically distinct from the 50 \kms cloud. Using the model described by \citet{Shull80} for a SNR in the snowplow phase and the mass estimate of the western streamer (molecular gas swept out of the central parsecs by Sgr A East), HH05 estimated the upper limit on the energy of the single progenitor supernova to be 9 $\times 10^{51}$ ergs. X-ray observations are consistent with a single supernova having a small gas mass and thermal energy of 10$^{49}$ ergs \citep{Maeda02}. The controversy has not been resolved, however, as \citet{Lee05} recently suggested that Sgr A East was created by a hypernova with an energy of 2-40 $\times 10^{52}$ ergs. 

Following the technique used in HH05 for the western streamer, the energy necessary for the Southern Runner to be accelerated by Sgr A East to its current velocity is 2 $\times 10^{51}$ ergs. Using this energy as the ``shell energy" in Equation (9) of HH05, we find the corresponding lower limit for the energy of the progenitor supernova to be $\sim$6 $\times 10^{51}$ ergs, which falls into the range of energies for Sgr A East observed by other studies. The value quoted in HH05 is an upper limit due to the fact that those authors performed further detailed analysis to determine the fraction of the SNR covered by the western streamer as well as to account for the expansion of the remnant into a non-homogenous environment; the energy quoted above does not take into account these details, and so it remains a lower limit.

To a first approximation, the average expansion velocity of Sgr A East can be determined from its current size (r $\sim$ 3.6~pc) and age ($\sim$ 10$^{4}$ years) to be of order a few $\times$ 100 km s$^{-1}$. The quoted age of Sgr A East is attributed to HH05 -- the energy estimate derived by Herrnstein \& Ho is   consistent with the estimate derived here, and so we assume their corresponding age for Sgr A East -- and to \citet{Maeda02}, wherein the same age is derived via the SNR's observed X-ray properties and the dynamical expansion model of \citet{Lozinskaya92} (which incidentally predicts a range of shock velocities for the blast wave down to 200 \kms). The velocity estimate is also consistent with the shock velocities of Sgr A East found by \citet{Lee05} to be $\sim$100 \kms \space from comparisons of H$_{2}$ and NH\3 line profiles. As the SNR is currently in its ``snowplow" phase in the dense environment of the Galactic center, this average expansion velocity is clearly an overestimate of the current motion, since the interstellar gas and dust swept up by the shock front would serve to slow the expansion during this phase. However, given the fact that our mass estimates are only reliable to an order of magnitude, a more complex determination of the kinematics of this SNR is not appropriate; the characteristic velocities observed are of the same order of magnitude as would be feasible for an expansion associated with Sgr A East and are consistent with the velocities found by other studies. 

The expansion of Sgr A East can also be estimated via velocity measurements of other features which are believed to be expanding with the edge of the remnant, much in the way that \citet{Zhao98} discuss the expansion of the ``eastern cavity" by measuring the proper motions of the H{\tiny \sc II} regions on its edge. \citet{Yusef99} observed OH masers assumed to be expanding with the edge of Sgr A East at velocities of +43 km s$^{-1}$ and +53 to +66 km s$^{-1}$. (A significant component of these velocities is perpendicular to the LOS.) \citet{McGary01} observed the northern ridge, a molecular feature believed to be expanding with the SNR, in NH\3 (3,3) at a velocity of --10 km s$^{-1}$ and an orientation also expected to be perpendicular to the LOS. Observations were also made of the western streamer, a filamentary molecular gas feature whose curvature, velocity gradient, line widths, and rotational temperatures are all consistent with the expansion -- again, highly inclined to the LOS -- of Sgr A East \citep{McGary01}. These examples illustrate the dependence of the analysis of the velocities in this region on the assumed geometry of the clouds. The high negative velocities of the NH\3 (6,6) are not inconsistent with those of other clouds assumed to be associated with Sgr A East, provided the NH\3 (6,6) is moving with a more significant component in the radial direction, parallel to our LOS.

Sgr A East is known to be expanding toward Sgr A* from behind, which supports the hypothesis that the high negative velocity NH\3 (6,6) presented here is expanding with it, as it has been shown that there exists much molecular gas -- including NH\3 (6,6) -- immediately surrounding Sgr A* (see \S4.3) that could have been accelerated to high velocities by the passage of this shock front. The velocity gradient of the Southern Runner, shown in Figure 4, span 40 km s$^{-1}$ in a space of $\approx$ 20$\as$ ($\sim$0.6~pc at the distance of the Galactic center), or 2 km s$^{-1}$ arcsec$^{-1}$. This potential explanation is further supported by the fact that the most rapidly moving gas in the Southern Runner is located closer to the projected center of Sgr A East. The slower moving gas is located closer to the edge (but still sufficiently far from it, as seen in Figure 5), which is consistent with a larger component of its velocity being perpendicular to the LOS. This particular feature may also be affected by the second supernova remnant expanding into Sgr A East from the southeast, G 359.92 -- 0.09 \citep{Coil00}. This second SNR is responsible for the concave southeastern edge of Sgr A East, highlighted by the green arc in Figure 5. The remnant has already been shown to affect other molecular features in the region, namely the southern streamer and the southern half of the molecular ridge as detected in NH\3 (1,1) and (2,2) by \citet{Coil00}. 

The same explanation may also be appropriate for the Cap, as the clouds slightly closer to the center (spectra B-D) have faster-moving components than the cloud toward the edge (spectrum A). This feature, however, is more complicated as there appear to be multiple velocity components represented in the closer clouds (spectra B-D) with less significant detections. If this feature is expanding with Sgr A East, the individual clouds may be exhibiting more chaotic motion relative to each other, causing their velocity components along our line of sight to vary, or perhaps there exist several clouds located at different positions along the LOS and thus moving with different velocities; observations achieving better signal-to-noise are clearly necessary in this case. The expansion of Sgr A East works well to explain the behavior of the Southern Runner, but it does not work as well for the Cap; it is likely that some other process is occurring at this location.

The association of the Southern Runner with the expansion of Sgr A East is further supported by the mass estimates of other low-velocity NH\3 clouds believed to have been swept up by the expansion of Sgr A East. The western streamer and northern ridge, observed in NH\3 (3,3) by HH05, are each estimated to be a few $\times$ 10$^{3}$ M$\sol$ with similar assumptions about X(NH\3) and temperature. This supports the feasibility of the idea that the high negative velocity emission was similarly ``swept up." Also, the clouds exhibiting the more excited transition may have been accelerated to higher velocities than their less excited counterparts. It is important to remember that there exist huge reservoirs of gas at the Galactic center in the form of GMC's (with masses of order $\sim$10$^{5}$ M$_{\odot}$) from which dense clouds of the sizes of the western streamer or northern ridge or those presented in this work could have been swept or tidally disrupted. These giant molecular clouds are observable in NH\3, as the 20 \kms \space GMC has been detected in NH\3 (1,1) and (2,2) \citep{Coil99,Coil00} and the 50 \kms \space GMC has been detected in NH\3 (3,3) \citep{McGary01,Herrnstein02} and are located within the central parsecs of the Galaxy. It is certainly possible that a shock, such as that of Sgr A East, pushing through the GMC's (or the molecular ridge connecting them) could strip smaller clouds of the sizes presented here, though observations achieving higher signal-to-noise are necessary to prove such a hypothesis. 

\subsection{The Circumnuclear Disk}

The circumnuclear disk, best observed in HCN (1-0), is shown in grayscale in Figure 6, again with an overlay of velocity integrated NH\3 (6,6) emission contours. As previously mentioned, one positional coincidence occurs at the location of the Cap, where the bright northern lobe of the CND is also found. However, the velocity of the HCN tracing the CND is +100 km s$^{-1}$ at this position (see Figs. 8 \& 9, \citealt{Wright01}), while the two velocity components of the NH\3 (6,6) are approximately --160 km s$^{-1}$ and --195 km s$^{-1}$. This coincidence may still be significant as the NH\3 (6,6) emission could be tracing a filament of gas falling from the CND toward Sgr A*. Again, as in the discussion of Sgr A East, if the NH\3 (6,6) emission is falling from the CND, its negative velocity indicates that the connecting gas would be falling toward Sgr A* from behind. 

If this filament hypothesis is true, and a gravitational or tidal interaction between the potential of the Galactic center and the Cap is occurring, then NH\3 (6,6) emission at all velocities between that of the northeastern lobe of the CND (+100 \kms) and those which we observe (--160 \kms) is missing. NH\3 (6,6) emission was not detected in the intervening velocity range at the position of the Cap by \citet{Herrnstein02}. The velocity resolution of that study was 9.3 \kms, and the rms noise in one channel was 8.7 mJy beam$^{-1}$. If a stream of gas does exist with a mass similar to that which we estimate for the Cap, this study would not have been sensitive to it; if located in the correct velocity range, the emission from the Cap would not have elicited a detection more significant than $\sim$1.5$\sigma$ in that study. (A more in depth discussion of the low-velocity NH\3 (6,6) emission that was detected by \citet{Herrnstein02} is discussed in the following section.) However, another possibility is that the gas falling from the CND is more clumpy in nature, or it was ``knocked" from the disk in a more shock-like event (such as the passage of the shock front of Sgr A East), in which cases extended filamentary structure would not be expected. This ``clumpy" nature is consistent with the line profile and position-velocity diagrams in that the emission is not continuous in velocity.

More recent studies of HCN(1-0) and HCO$^{+}$ by \citet{Christopher05} and \citet{Shukla04} estimate the total gas mass of the CND to be 10$^{6}$ M$\sol$ and 3 $\times$ 10$^{5}$ M$\sol$, respectively. \citet{Christopher05} estimate a typical dense gas core in the CND to be (2-3) $\times$ 10$^{4}$ M$\sol$, an order of magnitude greater than our mass estimate for the Cap. It is not surprising that the high negative velocity NH\3 (6,6) does not appear to belong to the same population as the gas cores of the CND. Instead, it is more likely that, if the features are physically associated, the hot gas was either shocked out of the ring or is falling from the CND toward the Galactic center under the strong potential created by Sgr A*.

Models have been created to predict the type of infall behavior that could be applicable to the Cap. One such model is described in \citet{Quinn85}. In this example, the authors recreate the structure of the minispiral with the infall and tidal breakup of a small molecular cloud that originated in or near the CND. The model filament, acting under the gravitational potential of the Galactic center and a dissipative medium, matches observations of the minispiral quite accurately, producing spectral features over a wide range of velocities. We propose that a similar situation could have created the Cap wherein a recent disturbance (the passage of the shock front of Sgr A East, for example) heated up a mass of molecular gas and decreased its volume density, rendering it susceptible to the tidal forces, thus ``knocking it out" of the CND from behind. The gravitational potential of the Galactic center would then continue to pull the cloud inward, though observations that could support the existence of such behavior have not been sensitive enough to prove it.

\subsection{Low-velocity NH\3 (6,6)}

Previous studies of NH\3 (6,6) at lower velocities (--140 km s$^{-1}$ to +130 km s$^{-1}$) found much emission within the central 2~pc of the Galactic center where lower transitions of the molecule were absent due to self-absorption along the LOS (HH05). In Figure 7, the high negative velocity NH\3 (6,6) is overlaid in contours on the low-velocity NH\3 (6,6) in grayscale. Position-velocity diagrams of low-velocity, high-line-ratio gas (defined by where the (6,6) flux is greater than the (3,3) flux; see Figure 13 of HH05) comprising the bright cloud in Figure 7 revealed that there exists a warm, dense cloud in the central 2~pc that appears to be rotating with a velocity of 20 \kms and is expanding at 80 \kms (HH05). This cloud is also coincident with a gap in the CND. As is evident in Figure 7, there is also more low-velocity NH\3 (6,6) emission farther from Sgr A* at distances similar to (and greater than) those of the Southern Runner and Cap. The high- and low-velocity observations of this transition may in fact still trace the same population of gas if the high-velocity NH\3 (6,6) is a small piece of the low-velocity, high-line-ratio gas that was stripped from the larger cloud by the Galactic potential, the expansion of Sgr A East, or winds from young stars as discussed in the introduction; however, we do not yet see any kinematic evidence for a physical connection. 

Performing the same calculation as shown in \S 3.1 on the low-velocity NH\3 (6,6) emission observed by HH05, we estimate 2 $\times$ 10$^{5}$ M$\sol$ for the bright, central, low-velocity (6,6) emission shown in grayscale in Figure 7. This estimate is substantially greater than the estimate for the high-velocity NH\3 (6,6) clouds and is therefore consistent with a model where the high-velocity gas was torn from the low-velocity gas.
\clearpage
\vspace{-0.5cm}
\begin{table}[t]
\small
\begin{center}
\centerline{Table~1: Observed Galactic Center Clouds}
\medskip
\begin{tabular}{lcccccc}
\hline
\hline
Tracer & Mass (M$\sol$) & $\Delta$v (km s$^{-1}$) & v$_{obs}$ (km s$^{-1}$) & $<$$\tau$(1,1)$>$ \\
\hline
Gas affected by Sgr A East$^{\it{a}}$ & (2-4)$\times$10$^{3}$ & 15 -- 20 & --70 to +90, --10 & 0.4 -- 0.5 \\ 
Low-velocity NH\3(6,6)$^{\it{b}}$ & 2$\times$10$^{5}$ & 20 & --20 to +100 & ... \\
HCN(1--0) emission $^{\it{c}}$ & (2-3)$\times$10$^{4}$ & 11 -- 40 & --110 to +110 & ... \\
OH absorption$^{\it{d}}$ &  5$\times$10$^{3}$ & 25 & --180 & 0.15 \\
CO emission $^{\it{e}}$ & 10$^{5}$ & 25 & --187 & ... \\
NH\3 (6,6) in EMR $^{\it{f}}$ & 6$\times$10$^{3}$ & 25 & 45, 160 & $\leq$0.5 \\
Southern Runner$^{\it{g}}$ & 4$\times 10^{3}$ & 15 & --165,--195 & ... \\
Cap$^{\it{g}}$ & 2$\times 10^{3}$ & 12 & --160,--195 & ... \\
20 km s$^{-1}$ GMC$^{\it{h}}$ & (2-3) $\times$ 10$^{5}$ & 30 -- 40 & 20 & 3.5 \\
\hline
\end{tabular}
\end{center}
\caption{Kinematic measurements and mass estimates for relevant molecular emission features at the Galactic center.}
\tablenotetext{a}{Refers to typical clouds pushed aside by Sgr A East, as discussed in text; specifically the western streamer and northern ridge as observed in (1,1), (2,2), and (3,3). (HH05)} 
\tablenotetext{b}{Refers to central cloud detected in NH\3 (6,6). \citealt{Herrnstein02, Herrnstein05}. Mass estimate using Equations (2) -- (4) and the assumptions detailed in this paper with T$_{b}$=81~K and $\Delta$v=85 \kms.}
\tablenotetext{c}{Refers to mass per dense gas core, as discussed in text. Total CND mass is $\approx$10$^{6}$ M$\sol$. \citealt{Christopher05}.} 
\tablenotetext{d}{\citealt{Zhao95}.} 
\tablenotetext{e}{\citealt{Liszt93}.}
\tablenotetext{f}{\citealt{Gardner87}.}
\tablenotetext{g}{This work; order of magnitude estimates assuming T$_{b}$=22.5~K.}
\tablenotetext{h}{\phantom{}Local quiescent GMC. Mass estimate from the southern cloud in \citet{Coil99,Zylka90}. Other parameters from \citet{Coil99}.}
\end{table}

\clearpage
\section{Conclusions}
Studying the temperature, morphology, and kinematics of the hot molecular gas at the Galactic center adds to our understanding of the complex nucleus of our own Galaxy and thus indirectly to our understanding of other galaxies. The detection of high negative velocity NH\3 (6,6) within 2~pc of Sgr A* leads us to conclude that any complete analysis of interactions at the Galactic center must include the effects of high velocity gas. 

Briefly, our results consist of the detection of NH\3 (6,6) in the velocity range of --142 to --210 \kms. Two distinct features appear: the Southern Runner, more coherent and continuous with velocity, and the clumpier Cap. The estimated total masses of these features are 4 $\times 10^{3}$ M$\sol$ and 2 $\times 10^{3}$ M$\sol$ respectively. Both features exhibit velocity structure at roughly --160 and --195 \kms. We propose several explanations for the clouds, assuming that they lie at the Galactic center: the Southern Runner could be expanding almost completely in the radial direction with the near edge of Sgr A East, or perhaps the Cap was rotating with the CND until the gravitational potential of the Galactic center or a shock front knocked the clouds from it and into infalling orbits toward Sgr A*. Both features could also have been ripped from a known low-velocity NH\3 (6,6) cloud by a similar disturbance or by stellar winds from young stars observed to be present near Sgr A*, as discussed in the introduction. Large reservoirs of gas of order 10$^{5}$ M$\sol$ are present in this region in the GMC's, so it is also not unlikely that clouds with these masses (as well as the northern ridge and western streamer) were once part of and then ripped from these larger clouds. 

As the line of sight orientation of these clouds relative to the other features at the Galactic center is very difficult to discern from these data, it is not currently possible to distinguish between these proposed explanations -- or other scenarios not presented here -- for the detected high velocity NH\3 (6,6) emission. (This is a common problem for studies in this region, e.g. \citealt{Gusten81, Zhao95, Coil00, Herrnstein05}).  However, dense gas accelerated to high velocities has been shown to exist at high temperatures in the central parsecs of the Galaxy, making it necessary to probe the entire velocity range further with higher signal-to-noise experiments in order to determine the spatial and velocity extents as well as the nature and origin of this high velocity gas. High negative velocity NH\3 (6,6) emission was also detected slightly outside of the FWHM of our primary beam, and though it was not discussed in this paper, it leads the authors to believe that more work must be done with this particular tracer throughout at least the central 10~pc. Combining interferometric data with single-dish data is also a necessary next step to understanding the spatial distribution and morphology of the high velocity gas. 

\acknowledgments
The authors wish to thank Jacqueline van Gorkom, David Helfand, Stephanie Tonnesen, and Ron Ekers for helpful comments. The authors also wish to thank the anonymous referee for improving this manuscript. This work was supported by the Astronomy Department at Columbia University. The authors also wish to credit the ADIL library for the image of Sgr A East used in Figure 5.


\clearpage
\begin{figure}
\includegraphics[width=5in]{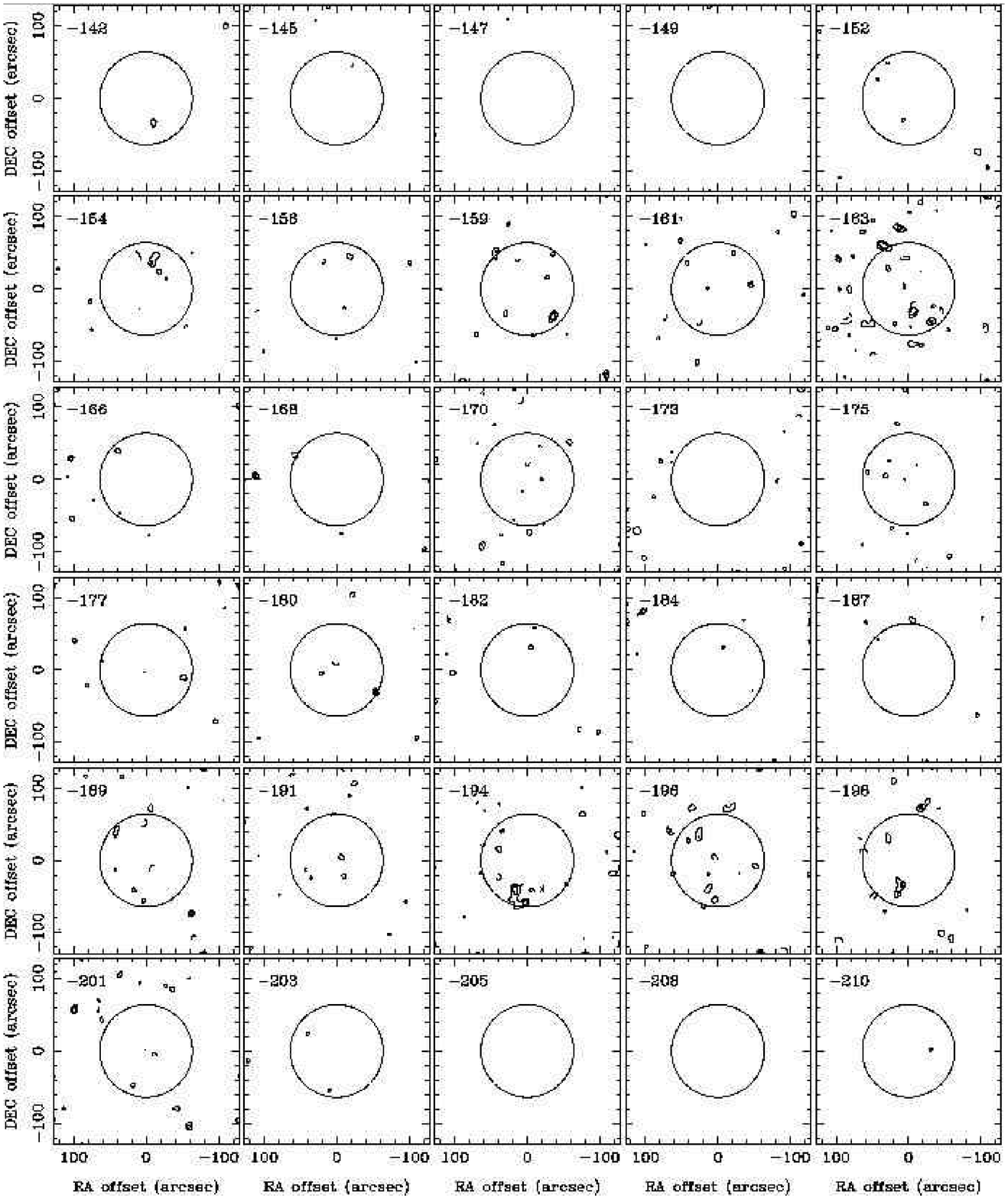}
\caption{Channel map of NH\3 (6,6) emission in 30 channels across the --142 to --210 \kms window. The overlaid circle, with a radius $\approx$ 2~pc, represents the FWHM of the primary beam for these data. Channels 8-11 and 23-25 (--159 to --166 and --194 to --198 \kms) have been box cleaned in order to suppress the negative sidelobes. The 1$\sigma$ noise level in each channel is $\approx$ 4 mJy beam$^{-1}$. The contours shown are 3.5$\sigma$ and 7$\sigma$. \label{1}}
\end{figure}

\begin{figure}
\includegraphics[angle=0,width=6.5in]{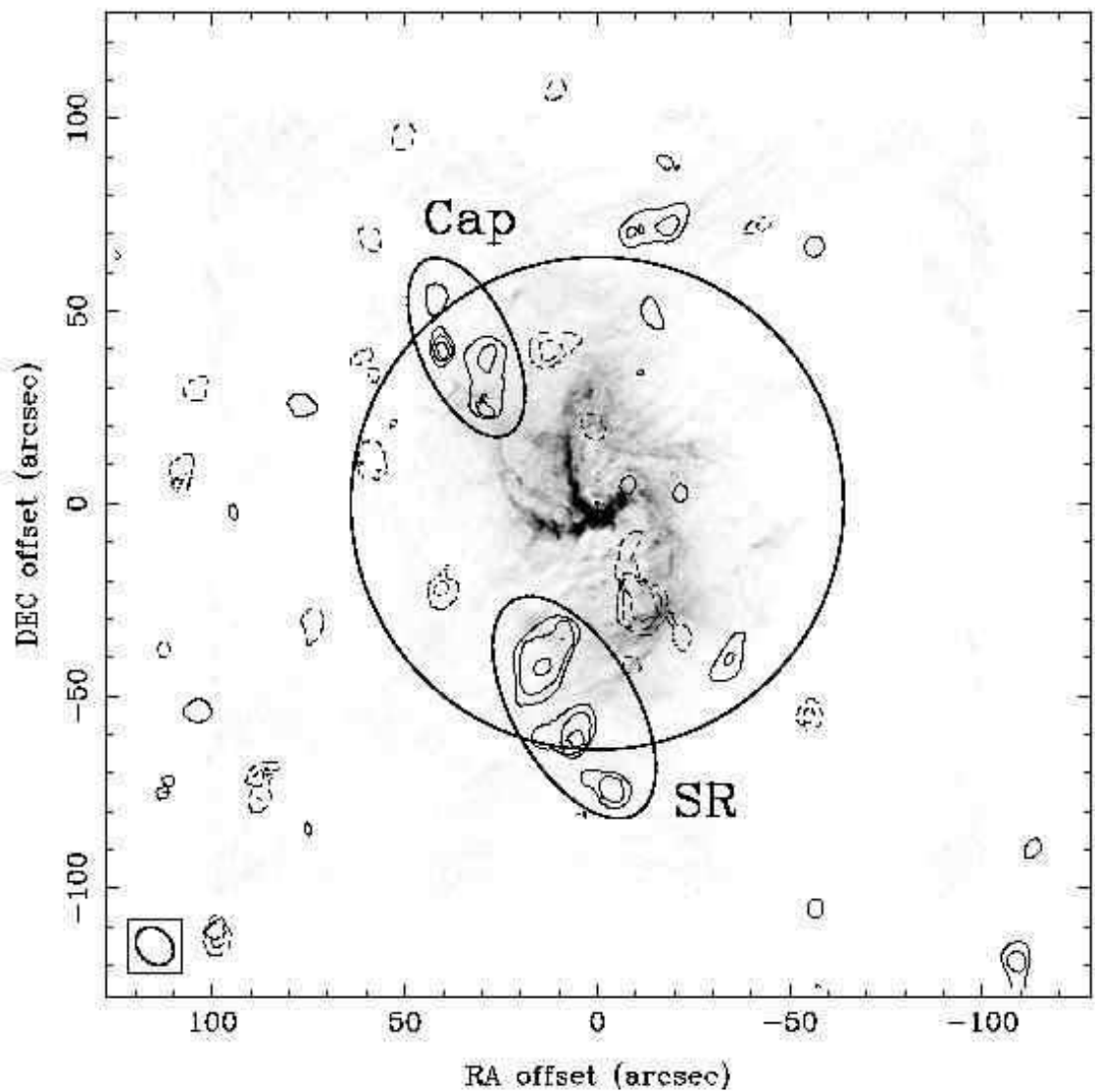}
\vspace{-4cm}
\caption{Velocity-integrated NH\3 (6,6) emission in the --142 to --210 km s$^{-1}$ range overlaid on continuum emission toward Sgr A West from \citet{Roberts93} in grayscale. The overlaid circle represents the FWHM of the primary beam, and the ellipses highlight the Southern Runner and Cap features. The (6,6) emission appears to be present beyond the extent of this primary beam. NH\3 (6,6) positive (solid line) and negative (dashed line) contours are in steps of 4$\sigma$, 6.5$\sigma$, and 9$\sigma$, where the typical 1$\sigma$ noise level is estimated to be 22.5 mJy beam$^{-1}$ \kms (see $\S$3). The beam size of the NH\3 (6,6) map is shown in the lower left corner for reference. The authors attribute the negative features to artifacts of the imaging process; they are not due to absorption. \label{2}}
\end{figure}

\begin{figure}
\includegraphics[angle=0,width=7in]{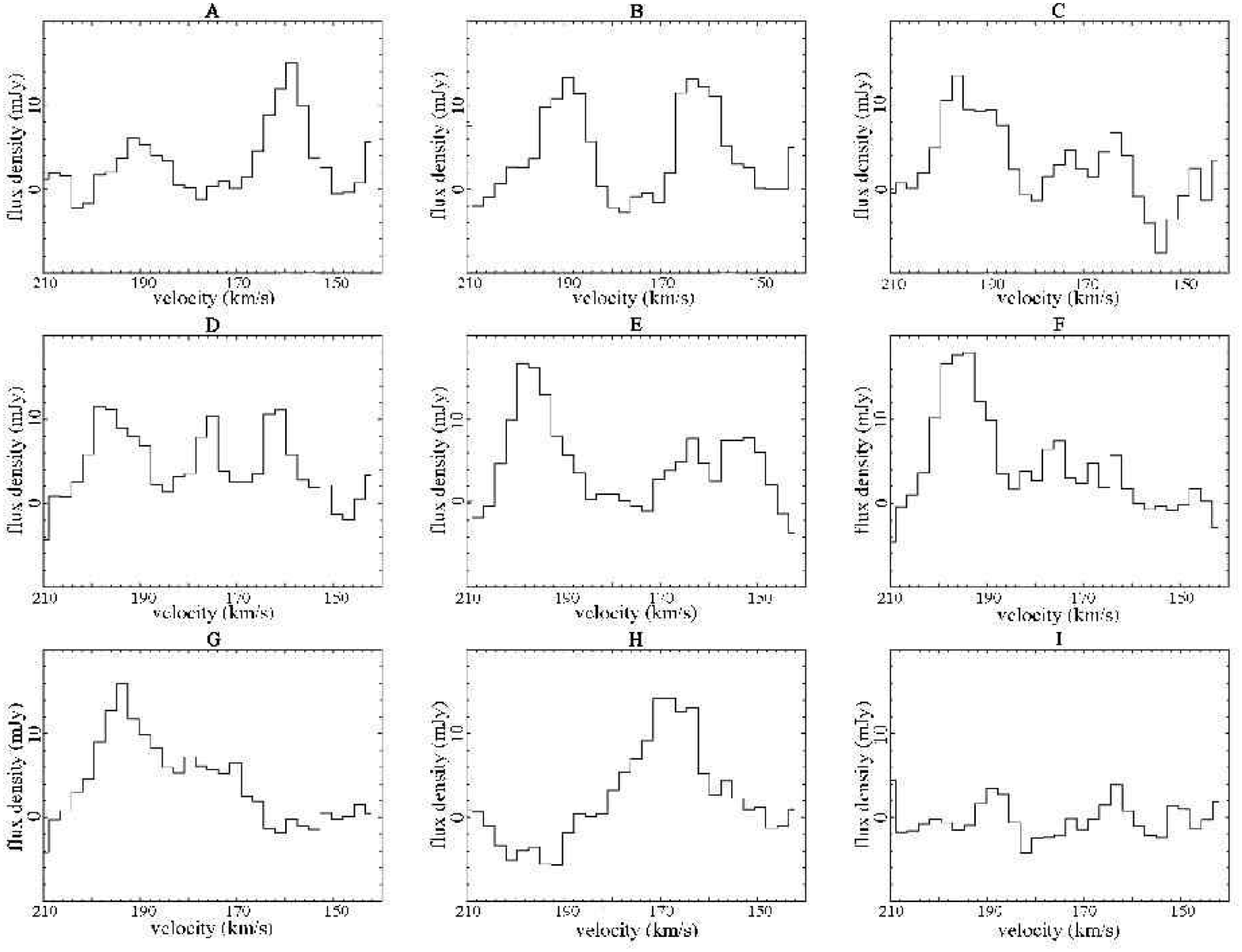}
\caption{Hanning-smoothed spectra taken at the positions specified in Figure 4. Spectra A-D are from the Cap, spectra E-H are from the Southern Runner, and spectrum I is included to give the level of the noise. The rms noise in any single channel is 4 mJy beam$^{-1}$. \label{4}}
\end{figure}

\begin{figure}
\begin{center}
\includegraphics[angle=0,width=3in]{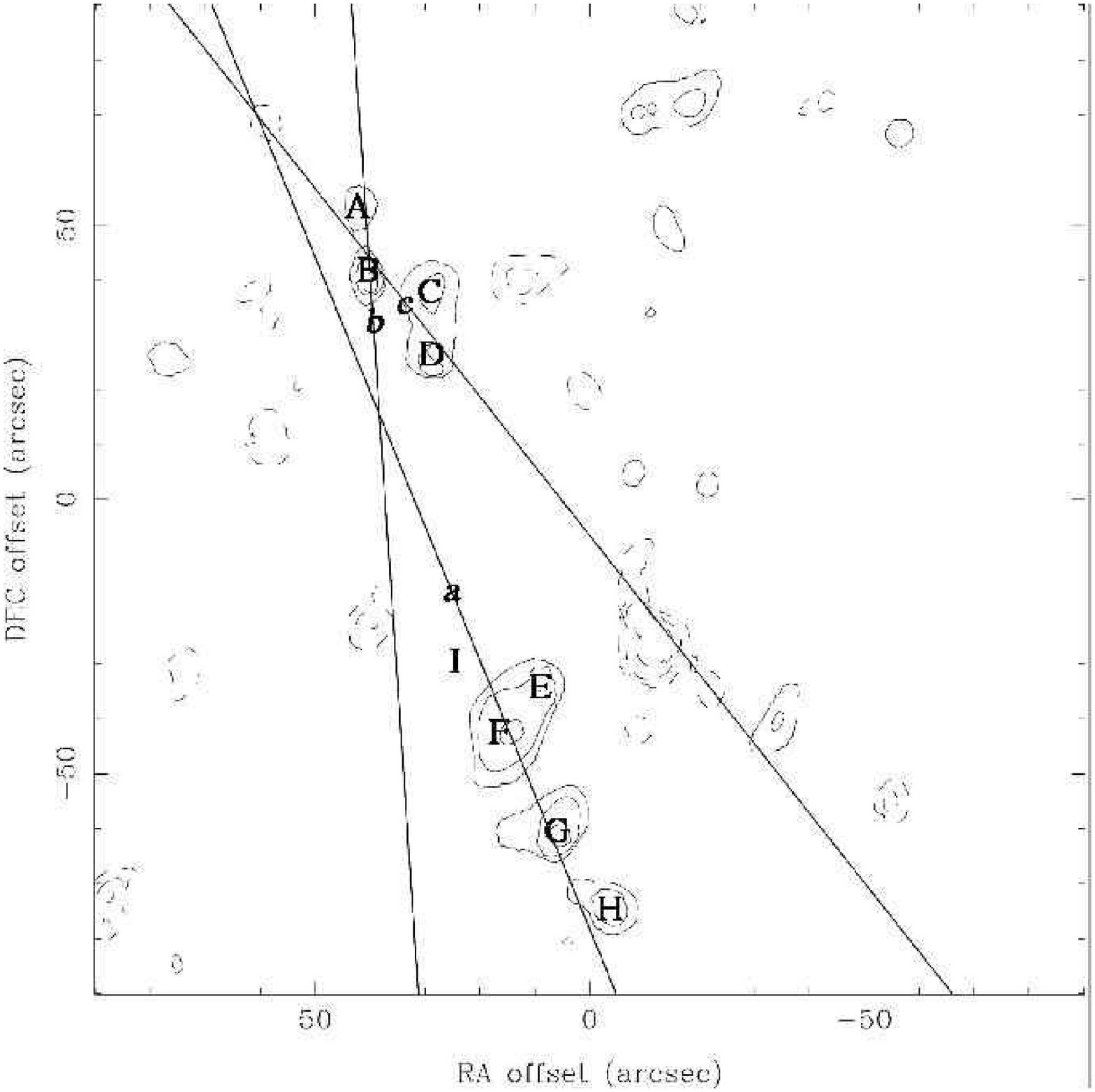}\hspace{10mm}
\includegraphics[angle=0,width=3in]{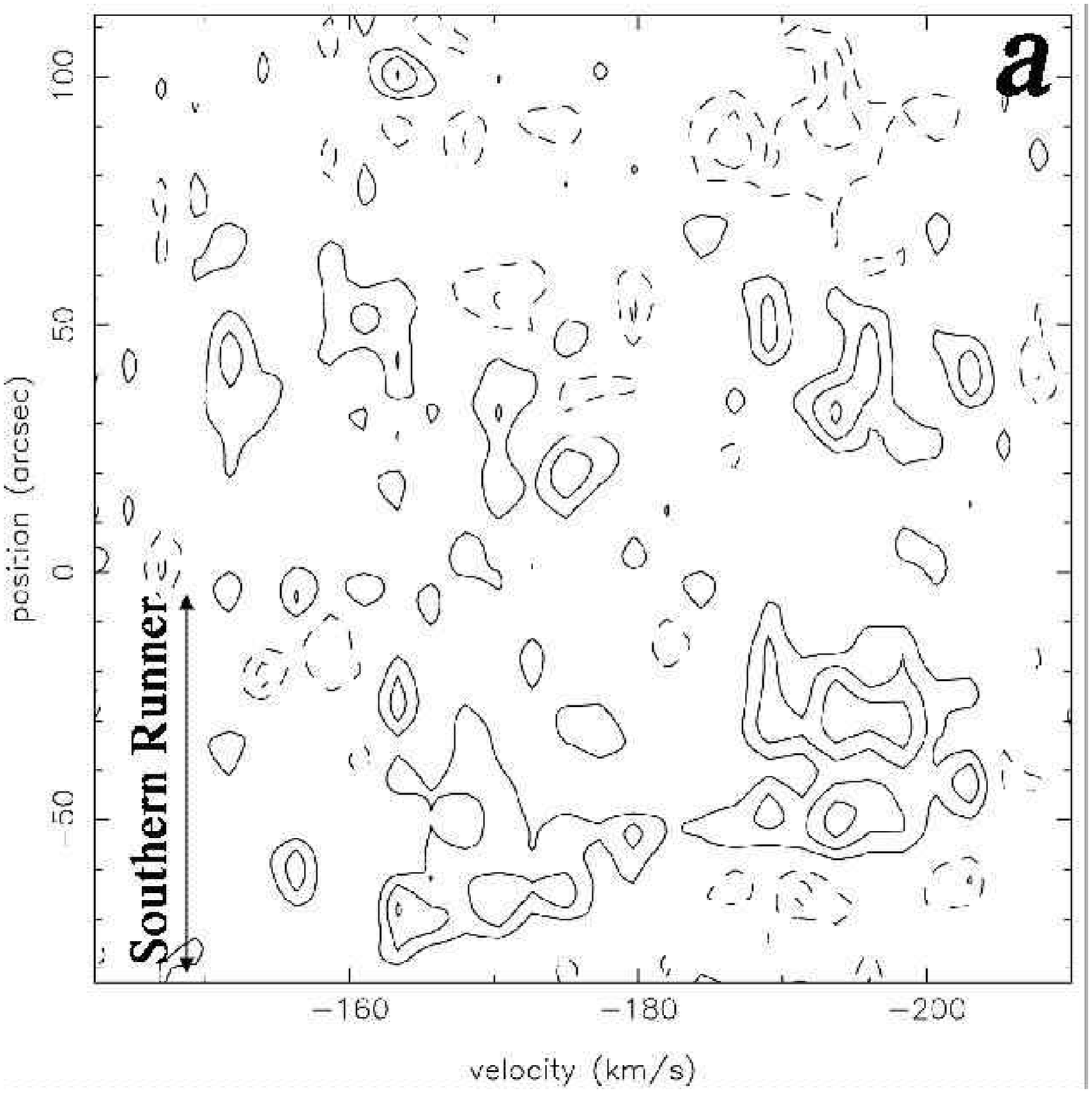}\\[10mm]
\includegraphics[angle=0,width=3in]{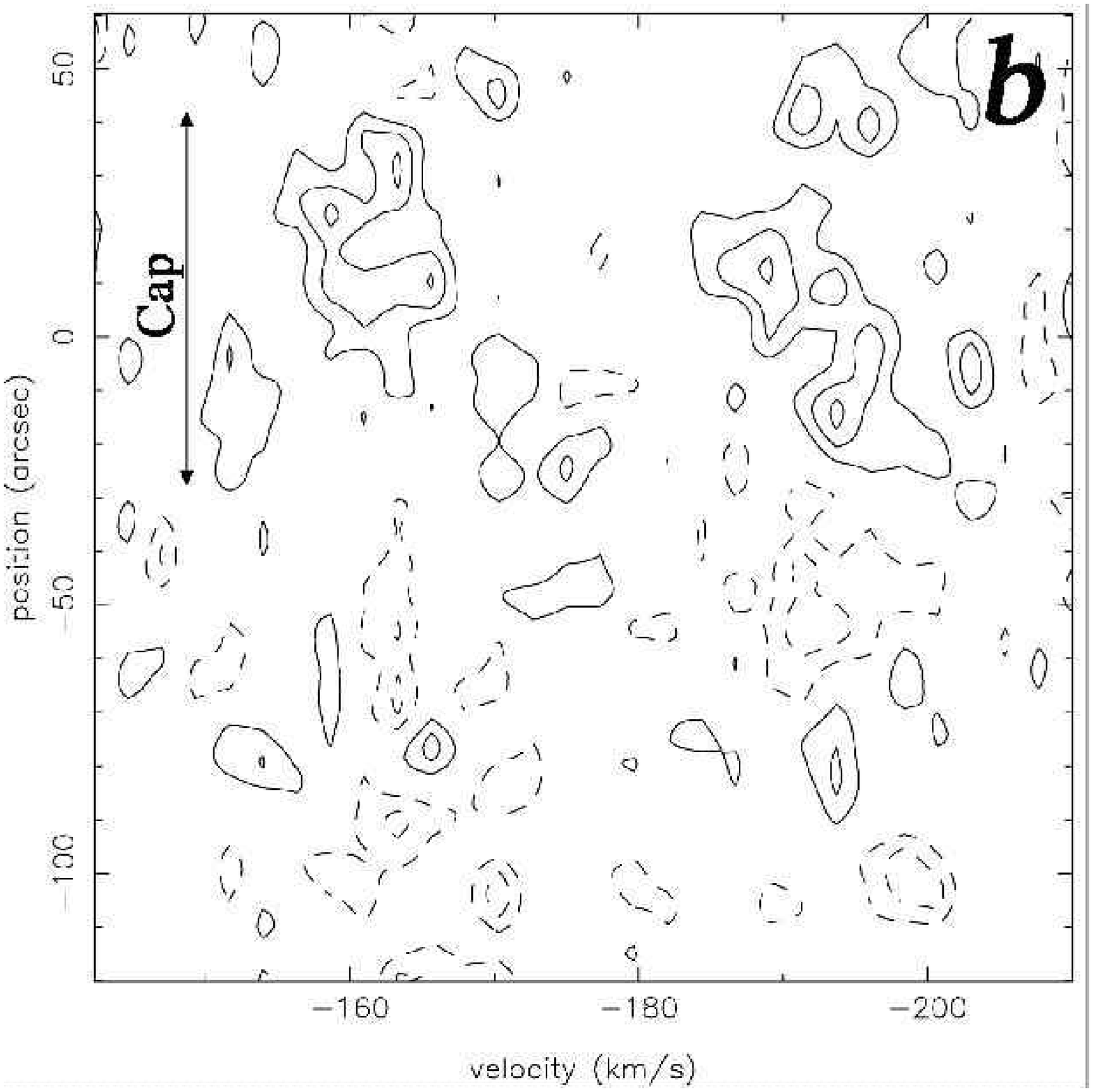}\hspace{10mm}
\includegraphics[angle=0,width=3in]{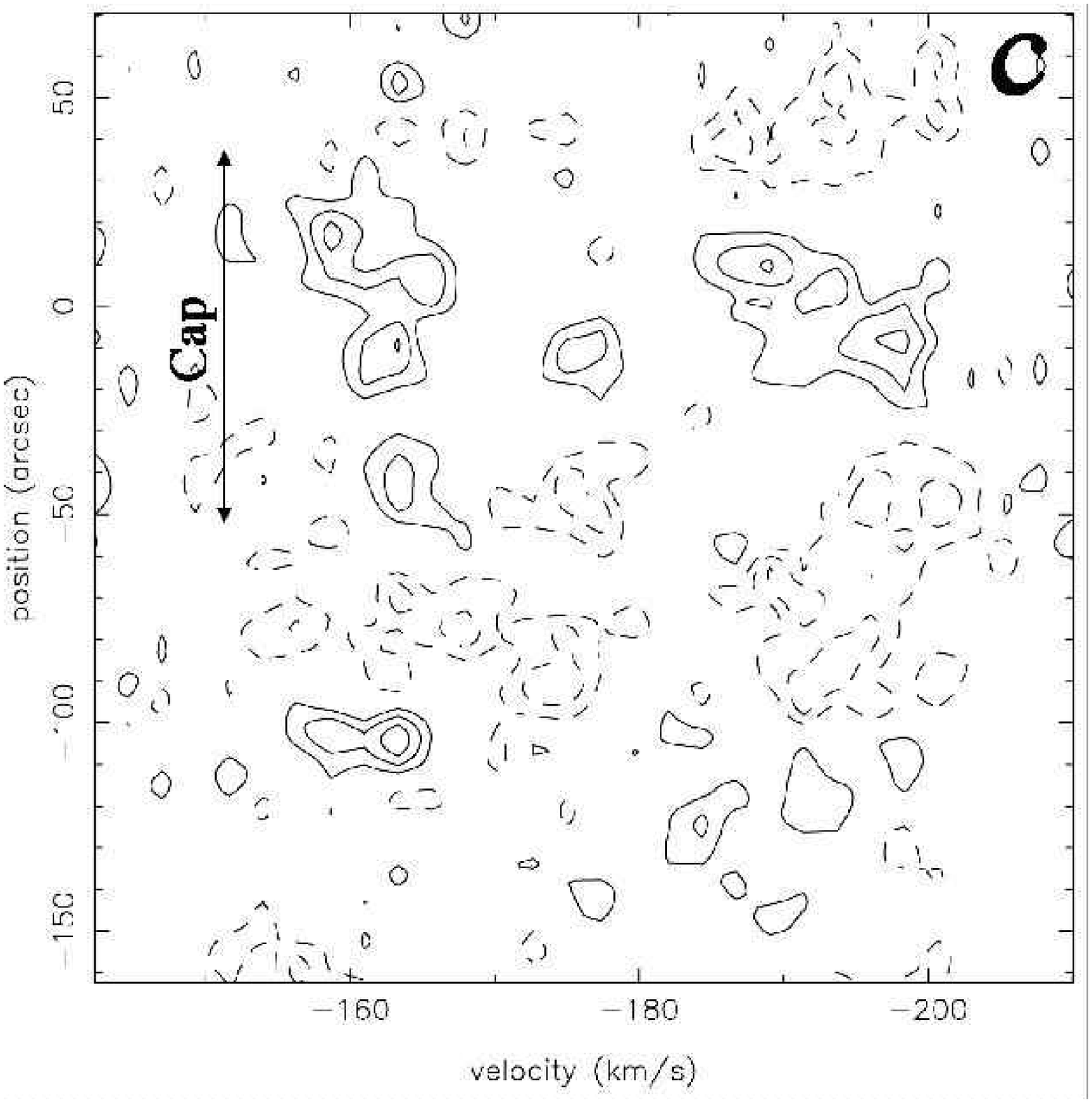}
\end{center}
\caption{Position-velocity diagrams along the specified cuts. The upper left map shows the velocity-integrated NH\3 (6,6) emission as shown in Figure 2 and displays the positions of the spectra in Figure 3. Sgr A* is located at (0,0). The upper right plot highlights the emission from the Southern Runner (cut a), while the lower left (cut b) and right (cut c) panels highlight the emission from the Cap. For reference, the approximate extent of the Southern Runner and Cap are labeled in each panel. The p-v diagrams are plotted with 1.25$\sigma$, 2.5$\sigma$, and 4$\sigma$ contours where 1$\sigma$ is the rms level in one channel, 4 mJy beam$^{-1}$. \label{5}}
\end{figure}

\begin{figure}
\includegraphics[angle=0,width=7in]{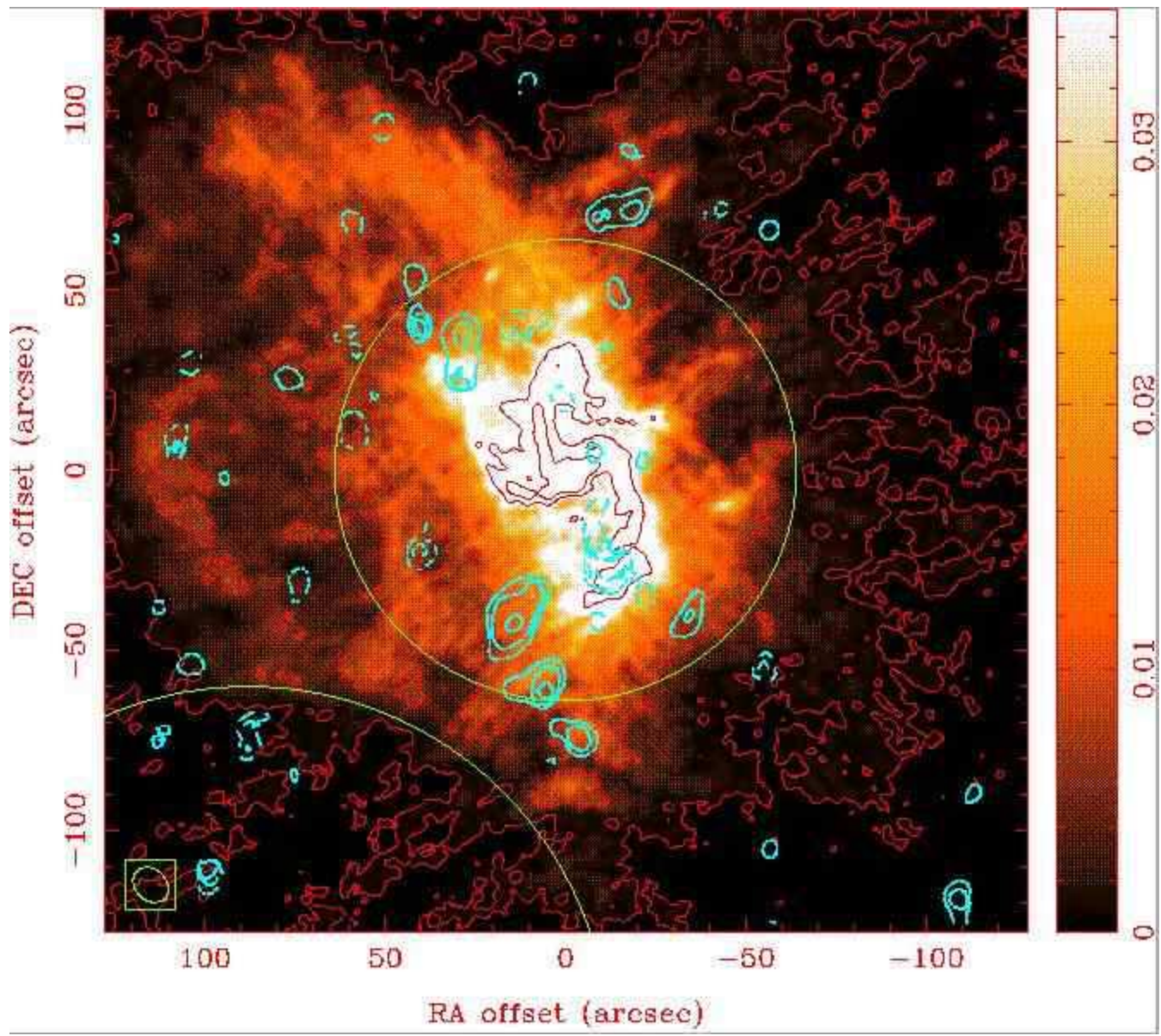}
\caption{Blue contours of NH\3 (6,6) overlaid on a color image of Sgr A East in 6 cm continuum \citep{Yusef87}. The red contours also measure the extent of the Sgr A East continuum, aiding the reader's eye through the excessive dynamic range of this emission. The concave southeastern edge of Sgr A East, highligted by the green curve, is due to the expansion of another supernova remnant (see $\S$4.1). The overlaid circle represents the FWHM of the primary beam for these data. The blue NH\3 (6,6) contours are in steps of 4$\sigma$, 6.5$\sigma$, and 9$\sigma$, where 1$\sigma$ = 22.5 mJy beam$^{-1}$ \kms, and the red contours are in steps of 1$\sigma$, 600$\sigma$, 5$\times$10$^{3}$$\sigma$, and 10$^{4}$$\sigma$ where 1$\sigma$ = 10$^{-5}$ Jy beam$^{-1}$. The beam size of the NH\3 (6,6) map is shown in the lower left corner for reference. \label{6}}
\end{figure}

\begin{figure}
\includegraphics[angle=0,width=7in]{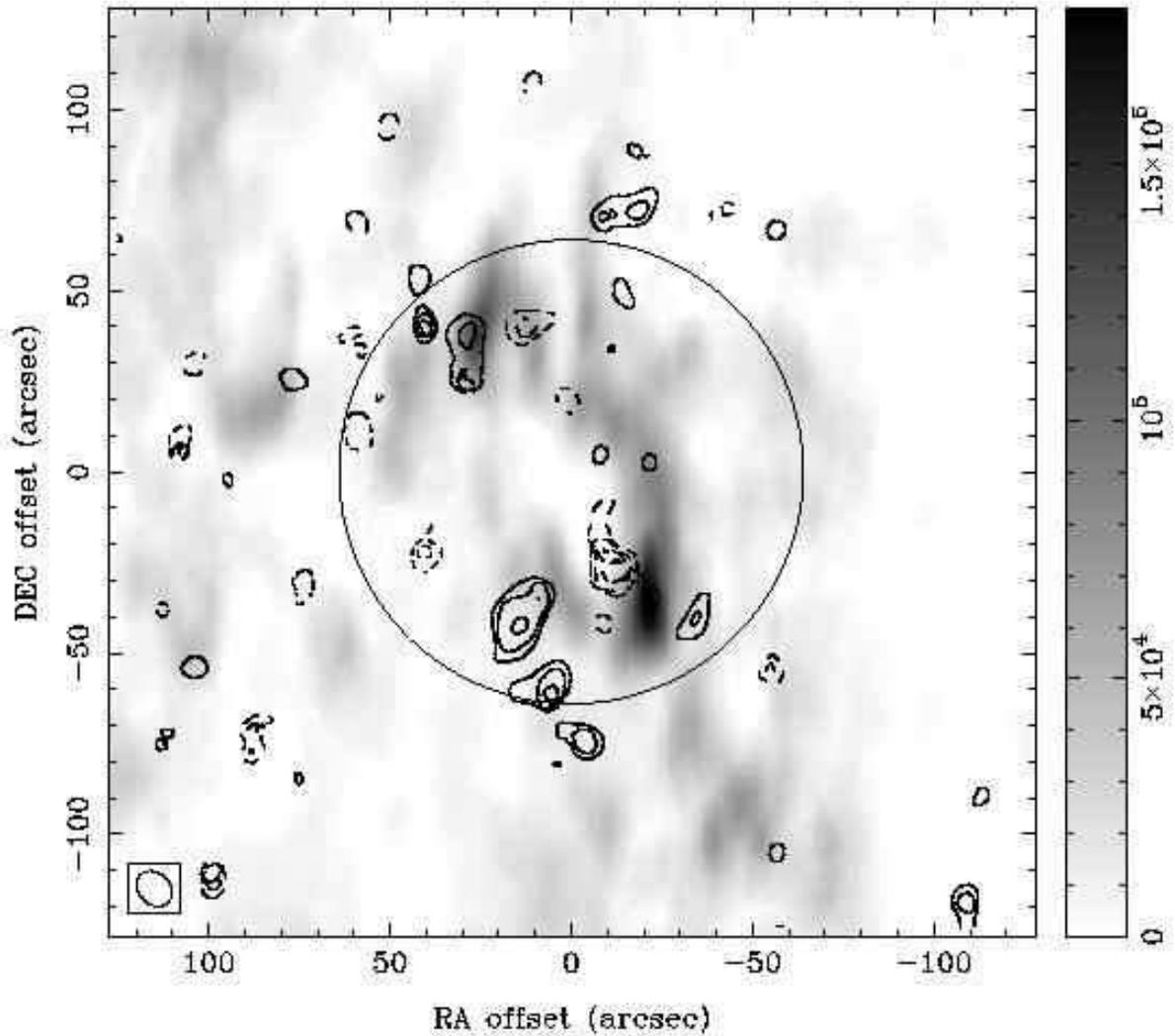}
\vspace{-5cm}
\caption{Contours of NH\3 (6,6) overlaid on a grayscale image of the CND in HCN (1-0) \citep{Wright01}. The overlaid circle represents the FWHM of the primary beam for these data. The NH\3 (6,6) contours are in steps of 4$\sigma$, 6.5$\sigma$, and 9$\sigma$, where 1$\sigma$ = 22.5 mJy beam$^{-1}$ \kms. The beam size of the NH\3 (6,6) map is shown in the lower left corner for reference.  \label{7}}
\end{figure}

\begin{figure}
\includegraphics[angle=0,width=7in]{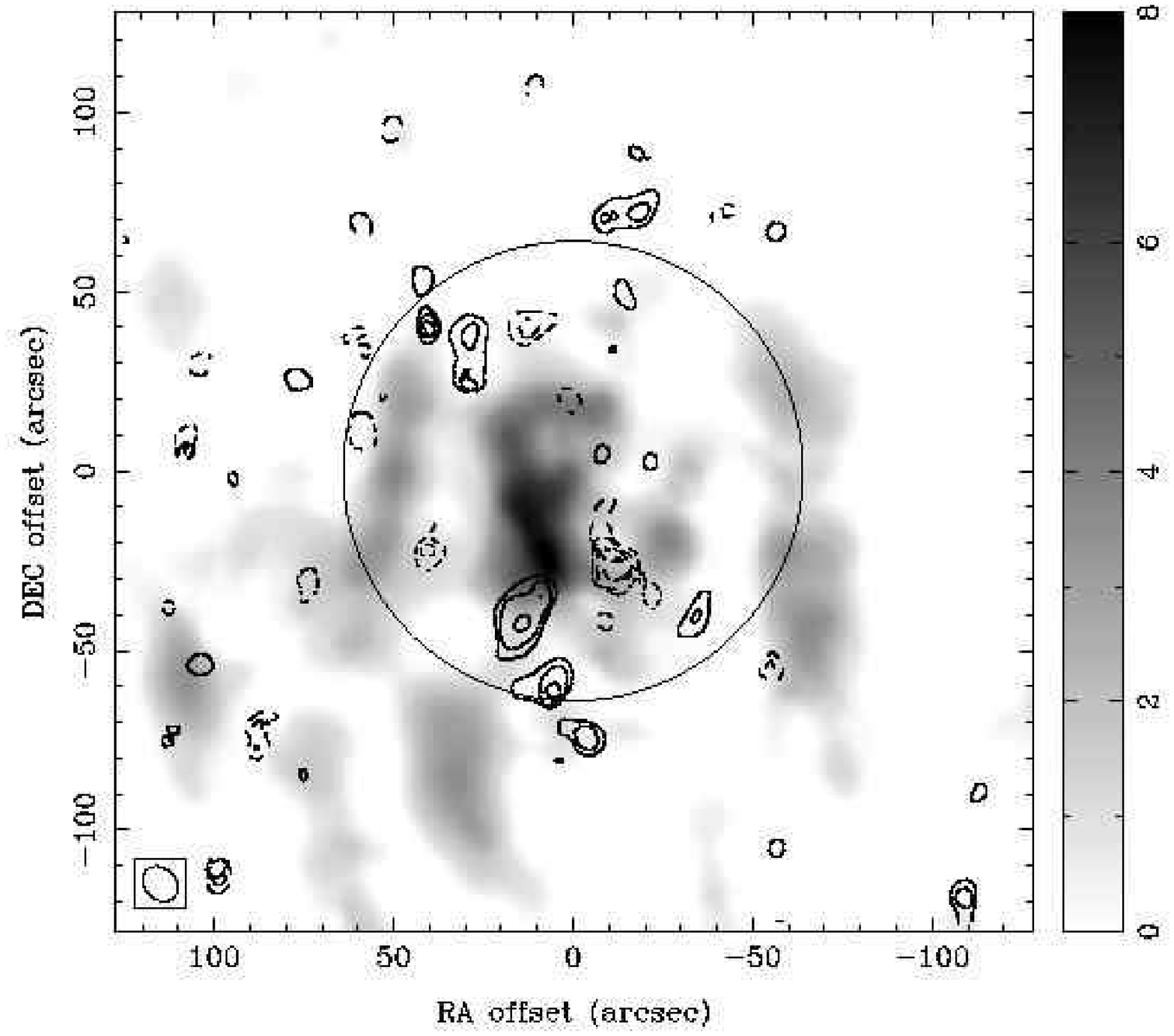}
\vspace{-5cm}
\caption{Contours of NH\3 (6,6) overlaid on a grayscale image of low-velocity (--140 to +130 \kms) NH\3 (6,6) observed by \citet{Herrnstein02}. The overlaid circle represents the FWHM of the primary beam for these data. The NH\3 (6,6) contours are in steps of 4$\sigma$, 6.5$\sigma$, and 9$\sigma$, where 1$\sigma$ = 22.5 mJy beam$^{-1}$ \kms. The beam size of the NH\3 (6,6) map is shown in the lower left corner for reference. \label{8}}
\end{figure}

\end{document}